\documentclass[conference]{IEEEtran}
\pdfoutput=1
\IEEEoverridecommandlockouts
\usepackage{cite}
\usepackage{amsmath,amssymb,amsfonts}
\usepackage{algorithmic}
\usepackage{graphicx}
\usepackage{textcomp}
\usepackage{xcolor}
\def\BibTeX{{\rm B\kern-.05em{\sc i\kern-.025em b}\kern-.08em
		T\kern-.1667em\lower.7ex\hbox{E}\kern-.125emX}}

\usepackage[english]{babel}
\usepackage[nolist]{acronym}

\begin{acronym}

\acro{3DES}{Triple-DES}

\acro{AES}{Advanced Encryption Standard}
\acro{ALU}{Arithmetic Logic Unit}
\acro{API}{Application Programming Interface}
\acro{ARX}{Addition Rotation XOR}
\acro{ASIC}{Application Specific Integrated Circuit}
\acro{ASIP}{Application Specific Instruction-Set Processor}
\acro{AS}{Active Serial}

\acro{BDD}{Binary Decision Diagram}
\acro{BGL}{Boost Graph Library}
\acro{BNF}{Backus-Naur Form}
\acro{BRAM}{Block-Ram}

\acro{CBC}{Cipher Block Chaining}
\acro{CFB}{Cipher Feedback Mode}
\acro{CFG}{Control Flow Graph}
\acro{CLB}{Configurable Logic Block}
\acro{CLI}{Command Line Interface}
\acro{COFF}{Common Object File Format}
\acro{CPA}{Correlation Power Analysis}
\acro{CPU}{Central Processing Unit}
\acro{CRC}{Cyclic Redundancy Check}
\acro{CTR}{Counter}

\acro{DC}{Direct Current}
\acro{DES}{Data Encryption Standard}
\acro{DFA}{Differential Frequency Analysis}
\acro{DFT}{Discrete Fourier Transform}
\acro{DLL}{Dynamic Link Library}
\acro{DMA}{Direct Memory Access}
\acro{DNF}{Disjunctive Normal Form}
\acro{DPA}{Differential Power Analysis}
\acro{DSO}{Digital Storage Oscilloscope}
\acro{DSP}{Digital Signal Processing}
\acro{DUT}{Design Under Test}

\acro{ECB}{Electronic Code Book}
\acro{ECC}{Elliptic Curve Cryptography}
\acro{EEPROM}{Electrically Erasable Programmable Read-only Memory}
\acro{EMA}{Electromagnetic Emanation}
\acro{EM}{electro-magnetic}

\acro{FFT}{Fast Fourier Transformation}
\acro{FF}{Flip Flop}
\acro{FI}{Fault Injection}
\acro{FIR}{Finite Impulse Response}
\acro{FPGA}{Field Programmable Gate Array}
\acro{FSM}{Finite State Machine}

\acro{GUI}{Graphical User Interface}

\acro{HDL}{Hardware Description Language}
\acro{HD}{Hamming Distance}
\acro{HF}{High Frequency}
\acro{HSM}{Hardware Security Module}
\acro{HW}{Hamming Weight}

\acro{IC}{Integrated Circuit}
\acro{IO}[I/O]{Input/Output}
\acro{IOB}{Input Output Block}
\acro{IOT}[IoT]{Internet of Things}
\acro{IP}{Intellectual Property}
\acro{ISA}{Instruction Set Architecture}
\acro{IV}{Initialization Vector}

\acro{JTAG}{Joint Test Action Group}

\acro{KAT}{Known Answer Test}

\acro{LFSR}{Linear Feedback Shift Register}
\acro{LSB}{Least Significant Bit}
\acro{LUT}{Look-Up Table}

\acro{MAC}{Message Authentication Code}
\acro{MIPS}{Microprocessor without Interlocked Pipeline Stages}
\acro{MMIO}{Memory Mapped \acl{IO}}
\acro{MSB}{Most Significant Bit}

\acro{NASA}{National Aeronautics and Space Administration}
\acro{NSA}{National Security Agency}
\acro{NVM}{Non-Volatile Memory}

\acro{OFB}{Output Feedback Mode}
\acro{OISC}{One Instruction Set Computer}
\acro{ORAM}{Oblivious Random Access Memory}
\acro{OS}{Operating System}

\acro{PAR}{Place-and-Route}
\acro{PCB}{Printed Circuit Board}
\acro{PC}{Personal Computer}
\acro{PS}{Passive Serial}
\acro{PUF}{Physical Unclonable Function}

\acro{RISC}{Reduced Instruction Set Computer}
\acro{RNG}{Random Number Generator}
\acro{ROM}{Read-Only Memory}
\acro{ROP}{Return-oriented Programming}
\acro{RTL}{Register Transfer Level}

\acro{SCA}{Side-Channel Analysis}
\acro{SHA}{Secure Hash Algorithm}
\acro{SNR}{Signal-to-Noise Ratio}
\acro{SPA}{Simple Power Analysis}
\acro{SPI}{Serial Peripheral Interface Bus}
\acro{SRAM}{Static Random Access Memory}

\acro{UART}{Universal Asynchronous Receiver Transmitter}
\acro{UHF}{Ultra-High Frequency}

\acro{VLSI}{very-large-scale integration}

\acro{WISC}{Writeable Instruction Set Computer}

\acro{XDL}{Xilinx Description Language}
\acro{XTS}{XEX-based Tweaked-codebook with ciphertext Stealing}

\end{acronym}

\usepackage[babel=true]{csquotes}
\usepackage[utf8]{inputenc}
\usepackage{array}
\usepackage{comment}
\usepackage[draft]{hyperref}
\usepackage{listings} 
\usepackage{makecell}
\usepackage{microtype}
\usepackage{tikz}
\usetikzlibrary{arrows}
\usetikzlibrary{automata}
\usetikzlibrary{circuits.logic.US}
\usetikzlibrary{decorations.text}
\usetikzlibrary{matrix}
\usetikzlibrary{shadows}
\usepackage{todonotes}
\usepackage{url}
\usepackage{xspace}

\newcommand{\HAL}{\textnormal{\textsf{HAL}}\xspace}

\definecolor{monokai_background}{RGB}{39, 40, 34}
\definecolor{monokai_blue}{RGB}{137, 189, 255}
\definecolor{monokai_green}{RGB}{166, 226, 42}
\definecolor{monokai_red}{RGB}{249, 38, 114}

\addto\extrasenglish{

}

\newcommand{\Figure}[1]{\textcolor{black}{\autoref*{#1}}}

\newcommand{\Section}[1]{\textcolor{black}{\autoref*{#1}}}
\newcommand{\Table}[1]{\textcolor{black}{\autoref*{#1}}}

\addto\captionsenglish{\renewcommand{\figurename}{Fig.}}
\begin{document}
	
	\title{Teaching Hardware Reverse Engineering: Educational Guidelines and Practical Insights\\
	}
	
	\author{\normalsize{Carina~Wiesen$^a$$^*$, Steffen~Becker$^b$$^*$, Marc~Fyrbiak$^b$, Nils~Albartus$^b$, Malte~Elson$^a$, Nikol~Rummel$^a$, and~Christof~Paar$^b$}\\
		\normalsize{$^a$\textit{Educational Research Institute} and $^b$\textit{Horst G\"ortz Institute for IT Security}}\\
		\normalsize{\textit{Ruhr-Universität Bochum}}\\
		\normalsize{Bochum, Germany}\\
		\normalsize{\{carina.wiesen, steffen.becker, marc.fyrbiak, nils.albartus, malte.elson, nikol.rummel, christof.paar\}@rub.de}\\\\
		\normalsize{\** Equal contribution}

		\thanks{This research was sponsored in part through ERC grant 695022 and NSF grant CNS-1563829.}
	}

	\maketitle

\begin{abstract}

Since underlying hardware components form the basis of trust in virtually any computing system, security failures in hardware pose a devastating threat to our daily lives. Hardware reverse engineering is commonly employed by security engineers in order to identify security vulnerabilities, to detect IP violations, or to conduct \ac{VLSI} failure analysis. Even though industry and the scientific community demand experts with expertise in hardware reverse engineering, there is a lack of educational offerings, and existing training is almost entirely unstructured and on the job.
To the best of our knowledge, we have developed the first course to systematically teach students hardware reverse engineering based on insights from the fields of educational research, cognitive science, and hardware security. 
The contribution of our work is threefold:
(1)~we propose underlying educational guidelines for practice-oriented courses which teach hardware reverse engineering;
(2)~we develop such a lab course with a special focus on gate-level netlist reverse engineering and provide the required tools to support it;
(3)~we conduct an educational evaluation of our pilot course. Based on our results, we provide valuable insights on the structure and content necessary to design and teach future courses on hardware reverse engineering.

\end{abstract}
	

\section{Introduction}
\label{ase:section:introduction}

In a world in which interconnected digital systems permeate almost all facets of our lives, IT security constitutes a major challenge for governments, individuals, and society at large. Even though various software layers employ a number of security mechanisms, hardware components are the basis of trust in virtually any computing system. Security failures at the hardware layer can have catastrophic consequences for the safety and security of computing systems as recently demonstrated by Spectre~\cite{CoRR:2018:Kocher} and Meltdown~\cite{CoRR:2018:Kocher:b}.

There are a number of reasons why inspecting digital hardware is highly desirable in a security context. First, malicious manipulations (e.g., hardware Trojans and backdoors) can compromise the security of an entire system~\cite{ches:2013:becker}. Since modern \ac{ASIC} design and fabrication processes are globally distributed, various (untrusted) stakeholders have access to valuable hardware designs and are thus able to commit piracy and perform malicious manipulations~\cite{ieee:2014:rostami}.
Second, low-quality counterfeits of integrated circuits pose serious security and safety risks, e.g., in mission-critical systems in aerospace or the power grid \cite{IEEE:2014:Guin, JET:2014:Guin}.
Third, it is estimated that semiconductor companies face losses in the range of several billion US dollars in global revenue due to hardware piracy~\cite{IEEE:2006:Pecht}. In this context, hardware reverse engineering experts play an essential role in helping companies detect violation of \ac{IP} in hardware designs.

The continuous evolution of a digital society shaped by a rapidly expanding \ac{IOT} and the proliferation of cyber-physical systems has created a high demand for IT security experts with a solid background in hardware reverse engineering. Nevertheless, there is an almost complete lack of educational offerings in the hardware reverse engineering field. Moreover, the topic of how to optimally structure an educational program that aims to teach hardware reverse engineering skills has not been extensively explored. Currently, hardware reverse engineering training happens almost entirely on the job and is restricted to a relatively small number of entities: Government agencies (for defensive and offensive purposes), large semiconductor companies (for competitive and failure analysis), and a small number of specialized hardware analysis companies. We argue that materially limited access to reverse engineering specialists leaves companies and institutions less able to identify and respond to hardware vulnerabilities, which in turn makes them more susceptible to exploitation and attack. We propose mitigating this industry-wide deficit of hardware reverse engineers through the integration of hardware reverse engineering training into existing security programs.

\par{\bfseries Goal and Contribution.}
In this paper, we motivate our work by sketching the high demand for experts in hardware reverse engineering and the surprising lack of educational courses. We then provide an introduction to the various elements essential to effectively teaching hardware reverse engineering which include technical knowledge, insights from the fields of educational research and cognitive science, and the utility of graphical representations, which play a crucial role in the field.

Based on this foundation, we propose educational guidelines for hardware reverse engineering courses, which we in turn used to create and conduct such a course. Finally, we present an educational evaluation of the pilot course, held in the winter term 2017/2018 at the Ruhr-Universität Bochum, Germany. In summary our contributions are:

\begin{itemize}
\item {\bfseries Educational Guidelines.}
To the best of our knowledge, we have designed the first structured guidelines to teach hardware reverse engineering with a particular focus on gate-level netlist reverse engineering.
Since the effective use of both graphical and textual representations is essential in the hardware reverse engineering field, our guidelines incorporate structures to support connection-making processes between both types of representations.
Our guidelines lay the foundation for a course structure that enables the acquisition of conceptual competencies such as sensemaking (e.g., skills to choose the \textit{meaningful} parts of a hardware design) or perceptional competencies (e.g., abilities to \textit{immediately} grasp the meaning of a graphical representation). These competencies support students' learning from graphical and textual representations, and their development in the context of learning hardware reverse engineering is consequentially essential.

\item {\bfseries Lab Course.}
Based on the proposed guidelines, we create a lab course consisting of five different projects with a special focus on gate-level netlist reverse engineering. We introduce our educational software environment based on the interactive tool HAL, which provides both textual and graphical representations of gate-level netlists.

\item {\bfseries Educational Evaluation.}
We provide valuable insights into the structure and content necessary to teach future courses in this area by conducting an educational evaluation, which considers perceived task difficulty, mental effort, and the level of relevant prior knowledge of the course participants. Based on the results, we derive methods for teaching and designing future courses on hardware reverse engineering. 
\end{itemize}


\section{Technical Background}
\label{ase:section:background}

The term \textit{reverse engineering} relates to the processes of extracting knowledge or design information from anything man-made in order to comprehend its inner structure~\cite{smc:1985:rekoff}. In the context of hardware security~\cite{book:hardware_obfuscation:chapter1}, security engineers (as well as adversaries) are forced to employ reverse engineering to make sense of a proprietary hardware design (e.g., to identify security vulnerabilities or security-circuitry for Trojan insertion~\cite{ivsw:2017:wallat}).
During this task, analysis of the gate-level netlist is a crucial step for human reverse engineers~\cite{ivsw:2017:fyrbiak}.

\par{\bfseries Gate-level Netlists.}
Synthesis tools convert \ac{RTL} descriptions of hardware designs into representations of the (Boolean) logic gates of the target gate library and their connectivity~\cite{Weste:2010:CVD:1841628}. Such representations are called gate-level netlists. A simple example of a gate-level netlist in (1)~graphical representation and the equivalent (2)~textual representation can be found in \Figure{ase:figure:synthesis_example}.

During the different synthesis steps, valuable high-level information such as (1)~meaningful descriptive information (e.g., names and comments), (2)~boundaries of implemented modules, and (3)~module hierarchies is lost. In practice, this loss of information highly complicates the reverse engineering process~\cite{tect:2013:subramanyan}.

In real-world settings, analysts can obtain gate-level netlists in several scenarios: (1)~through chip-level reverse engineering  in the case of a given \ac{ASIC} (involving steps such as decapsulation and delayering)~\cite{ches:2009:torrance}, (2)~through bitstream reverse engineering in the case of \acp{FPGA}~\cite{fpga:2008:note}, or (3)~by direct access at a foundry or through bribery or theft.

\par{\bfseries Tools.}
Similar to complex hardware design processes, hardware reverse engineering requires tools to automate time-consuming tasks and simplify steps for human analysts (e.g., through different representation forms). In particular, the latter is important for teaching this topic to novices in this area. Even though several tools exist in the industrial sector~\cite{ches:2009:torrance,url:ares}, such programs are typically not publicly available. In anticipation of \Section{ase:section:environment}, we selected the hardware reverse engineering tool \HAL~\cite{tdsc:2018:fyrbiak} as our educational software environment since it provides a rich-featured interactive \ac{GUI} suited for manual analysis, graph-based visual representations of gate-level netlists, and built-in extensibility for the integration of custom functionalities.

\par{\bfseries Representations.}
Various representational forms are involved during hardware design processes as well as reverse engineering processes. For example, during design and implementation phases, simulation waveforms are typically analyzed for debugging purposes. During gate-level netlist reverse engineering, textual forms or graph-based representational forms are analyzed, see \Figure{ase:figure:synthesis_example}. Therefore, the process of hardware reverse engineering necessitates the utilization of graphical representations.

\begin{figure*}[!htb]
	\centering

\begin{tikzpicture}[circuit logic US]
    
	\tikzstyle{every state}=[fill=white,draw=black,text=black,inner sep=0pt,minimum size=17pt,circular drop shadow]

	\tikzstyle{every circuit symbol}=[fill=white,drop shadow,scale=0.75]

	\draw [help lines] (0,0) (14,6);

	\node[initial, initial text={\scriptsize RST}, state, circle split] (s0) at (2,4) {\footnotesize $S_0$ \nodepart{lower} \footnotesize 1};
	\node[state, circle split] (s1) at (4,4) {\footnotesize $S_1$ \nodepart{lower} \footnotesize 0};

	\path[->,>=stealth'] 
		(s0) edge [loop above] node {\small0} (s0)
		(s1) edge [loop above] node {\small0} (s1)
		(s0) edge [bend left, anchor=center, above] node {\small1} (s1)
		(s1) edge [bend left,anchor=center, below] node {\small1} (s0);

    
    \begin{scope}[yshift=-0.65cm]
	\node [not gate, inputs={n}]  (not) at (4.5,2) {G3};
	\node [xor gate, inputs={nn}] (xor) at (1.5,2) {G1};

	\draw[drop shadow, fill=white] (2.5, 1.25) rectangle (3.5, 2.25) node at (3,1.75) {\footnotesize G2};
	\draw (2.5, 1.25 + 0.15) -- (2.5 + 0.15, 1.25 + 0.15 + 0.075) -- (2.5, 1.25 + 0.15 + 0.15);

	\draw [->,>=stealth'] (3.5, 2) -- (not.input);
	\draw [->,>=stealth'] (4, 2) -- (4, 3) -- (0.5, 3) -- (0.5, 2.075) -- (xor.input 1);
    \draw [fill=black] (4,2) circle [radius=1pt];
	\draw [->,>=stealth'] (not.output) -- ([xshift=+0.5cm] not.output) node at ([xshift=+0.75cm]not.output) {\small O};
	\draw [->,>=stealth'] ([xshift=-0.715cm]xor.input 2) -- (xor.input 2) node at ([xshift=-1cm]xor.input 2) {\small I};
	\draw [->,>=stealth'] (xor.output) -- (2.5, 2);
    \end{scope}

    \draw [fill=monokai_background,draw=white,rounded corners] (6.5,0) rectangle (13.5,5.75);

	\node at (10, 3) {
	\begin{lstlisting}[
    	basicstyle=\sffamily\footnotesize\color{white},
    	keywordstyle=\ttfamily\bfseries\color{monokai_blue},
    	stringstyle=\bfseries\color{yellow},
    	commentstyle=\bfseries\color{gray},
        escapechar=@,
    	morekeywords={module,endmodule, input, output, wire},
        literate={,}{{\textcolor{monokai_red}{,}}}{1}
            {;}{{\textcolor{monokai_red}{;}}}{1}
            {(}{{\textcolor{monokai_red}{(}}}{1}
            {)}{{\textcolor{monokai_red}{)}}}{1}
    ]
module FSM (RST, CLK, I, O);
    input RST, CLK, I;
    output O;
    wire o_G1, o_G2;
    @\textcolor{monokai_green}{XOR G1}@ (
    	.IN1(o_G2), .IN2(I), .O(o_G1)
   	);   
    @\textcolor{monokai_green}{DFF G2}@ (
    	.CLK(CLK), .RST(RST), 
        .D(o_G1),  .Q(o_G2)
    );
    @\textcolor{monokai_green}{INV G3}@ (
    	.IN(o_G2), .O(O)
    );
endmodule
	\end{lstlisting}
	};
\end{tikzpicture}
	\caption{Example Moore \acf{FSM} circuit as state transition graph (upper left) with associated gate-level netlist in (1)~visual graph-based representation (lower 
		left), and (2)~textual representation with an exemplary gate library in Verilog (right).}
	\label{ase:figure:synthesis_example}
\end{figure*}

\section{Educational Guidelines}
\label{ase:section:guidelines}

In order to derive guidelines for the structure of courses which allow effective teaching of hardware reverse engineering, we first summarize relevant current background in educational research and cognitive science regarding learning with graphical and textual representations. 

\subsection{Pedagogical Background}

Both textual and graphical representations play a central role in hardware reverse engineering (see~\ref{ase:section:background}). From a pedagogical point of view, two major challenges have to be solved to teach hardware reverse engineering effectively: (1)~Students need support in learning from and working with domain-specific graphical representations, and (2)~students' connection-making abilities between textual and graphical representations need to be facilitated. To this end, we analyze relevant current background in educational research, which provides guidance as to how these two challenges can be met. Note that the cognitive theory of multimedia learning~\cite{2005:mayer, 2009:mayer}, and the integrated model of text and picture comprehension~\cite{2005:schnotz} distinguish between learning from graphical and from textual representations. This distinction is based on the processing of text in the verbal part of the working memory~\cite{1992:baddeley,2012:baddeley}, and the processing of graphical representations in the visual part~\cite{1992:baddeley,2012:baddeley}. Thus, these different processing pathways cause different demands for an educational course structure, in which graphical and textual representations are combined. 

\subsubsection{Opportunities and Challenges -- Learning with Graphical Representations}

Prior educational research outlined opportunities for as well as challenges with learning from graphical representations. Schnotz~\cite{2014:schnotz} showed that graphical representations can support students' learning success by making abstract concepts more accessible. Additionally, graphical representations can depict supplementary information~\cite{2006:ainsworth,2014:ainsworth} which enables students to build a deeper understanding of novel content~\cite{2003:seufert}. Sociocultural theories consider graphical representations as an important form of scientific communication~\cite{2000:kozma,2014:wertsch}.
Despite their value in supporting students' learning processes, graphical representations can also be challenging for students. One of the main challenges in this context is the \textit{representational dilemma}~\cite{2006:ainsworth,2015:mcelhaney,2017:rau}. A representational dilemma exists when students have to learn new content knowledge they do not yet understand from graphical representations they also do not yet understand~\cite{2017:rau}. To overcome this challenge and benefit from graphical representations, it is necessary for students to develop specific competencies. These competencies support students in recognizing how graphical representations depict relevant information in order to solve a task or to learn new content. Thus, it is essential to consider how the development of specific competencies can be supported by the structure of the course. In the following, the specific competencies to overcome the representational dilemma and to benefit from domain-specific graphical representations will be addressed. Based on educational research, we outline how these competencies can be acquired to reach the two goals of (1)~enabling students' learning from and working with graphical representations, and (2)~facilitating students' connection-making abilities between textual and graphical representations.

\subsubsection{Competencies for Learning from and Working with Graphical Representations}

According to cognitive learning theories, students have to acquire representational competencies to overcome the representational dilemma~\cite{2005:gilbert,2008:gilbert,Ainsworth2008,1998:dejong,2015:mcelhaney}. Representational competencies are defined as the skills and knowledge applied to interpret and use graphical representations~\cite{2005:sim}. Cognitive learning theories distinguish two types of representational competencies: (1)~conceptual competencies, and (2)~perceptual competencies. Prior findings showed that these two types of representational competencies are linked to learning processes with graphical representations~\cite{2005:mayer,2009:mayer,2005:schnotz,2014:schnotz,2017:rau} and are consequently relevant for educational guidelines.

\par{\bfseries Conceptual Competencies: Sensemaking.}
Conceptual competencies describe a set of skills and practices that enable students to relate graphical representations to prior knowledge, to draw inferences based on graphical representations, and to choose the most suitable graphical representation which contains the information necessary to complete a task~\cite{2008:michalchik,2017:rau}. The process of choosing the relevant parts of a graphical representation for solving a problem depends on the students' ability to identify meaningful visual features. Koedinger et al.~\cite{2012:koedinger} showed that students' development of conceptual competencies is based on sensemaking processes. Dougherty et al.~\cite{2006:dougherty} described sensemaking as a process in which a person combines various information and ideas in a meaningful way. Sensemaking processes are verbally mediated~\cite{1989:chi,1983:gentner,2012:koedinger,2017:rau} and explicit since students need to willfully engage in verbal explanations~\cite{2000:disessa}.

\par{\bfseries Perceptual Competencies: Fluency.}
Perceptual competencies are defined as the ability to immediately detect the meaning of a graphical representation. The immediate detection of relevant information from a graphical representation is accomplished effortlessly and efficiently through an improved ability to recognize visual patterns~\cite{1996:gibson,2005:mayer,2009:mayer,2005:schnotz,2014:schnotz}. Perceptual competencies include the concept of fluency in recognizing and processing information from and about the graphical representation~\cite{2010:rocke}. The acquisition of perceptual competencies does not require direct instruction, but rather experience-based learning through the repetition of numerous examples~\cite{1996:gibson,2000:gibson,2013:kellman,1996:richman}.

\subsubsection{Competencies for Connection-Making Abilities}

Besides representational competencies, which are necessary to benefit from graphical representations, students have to acquire skills and knowledge about how to glean information from textual and graphical representations in tandem. The main phenomena here are the ways in which the graphical representation constrains the understanding of the text as well as the mechanisms by which text and graphics complement each other to convey relevant information~\cite{2006:ainsworth,2014:ainsworth}. Acquiring competencies in connection-making between the textual and graphical representations is essential; since both types of representation depict relevant information in gate-level netlists (see~\Figure{ase:figure:synthesis_example}). Schnotz and Bannert~\cite{2003:schnotz} outlined that text and picture comprehension are goal-oriented processes, in which the individual actively selects and handles verbal and graphical information to construct mental representations to complete the task at hand. The general assumption that adding graphics to a text improves the learning process overlooks the fact that graphical representations can be redundant to the accompanying text or dependent on the level of prior knowledge of the learners. Prior findings showed that learners with low prior knowledge benefit from combining textual and graphical representations, whereas learners with high prior knowledge also seem to be able to learn content from text alone~\cite{2003:schnotz}. The learning processes for acquiring connection-making abilities are described as verbally mediated sensemaking processes that use graphical representations in authentic tasks~\cite{2017:rau}.

In summary, hardware reverse engineering processes are greatly facilitated by working with textual and graphical representations of gate-level netlists. Consequently, a course imparting content from and skills in this field has to integrate these two forms of representations. Prior educational research has shown that learning with graphical representations can be beneficial and challenging at the same time. To overcome these challenges, students have to acquire specific competencies. These can be acquired in different ways, which have to be considered while designing content and structure of a new course. Based on these demands we propose four guidelines for an educational course which teaches hardware reverse engineering.

\subsection{Educational Guidelines for Teaching Hardware Reverse Engineering}
\label{ase:section:guidelines:hwreg}

\par{\bfseries (1) Integration of Graphical Representations.}
By drawing upon educational research, we have summarized how graphical representations can enable access to complex concepts~\cite{2014:schnotz} and support students' learning processes. Since working with graphical representations is a common practice in hardware reverse engineering, graphical representations should be an integral part of a course which teaches reverse engineering.

\par{\bfseries (2) Instructional Support to Develop Conceptual Competencies.}
Working with graphical representations can be challenging for students, especially when they are using representations for the first time. To overcome the representational dilemma, special instructional support is needed~\cite{2017:rau} to acquire conceptual competencies based on the activation of sensemaking processes~\cite{2012:koedinger}. Consequently, instructional support to acquire conceptual competencies needs to engage students in active reasoning, for example by prompting students to self-explain how they used a graphical representation or by engaging in discussions with other students about solving a task~\cite{2015:fiorella,2012:koedinger,2017:rau}.

\par{\bfseries (3) Support to Develop Perceptual Competencies.}
When integrating graphical representations into a course it is important to support the development of students' fluency-ability to recognize the meaning of a graphical representation immediately.
Based on prior findings, the development of perceptual competencies is not based on instructional support but on experience-based learning with repetition of numerous examples~\cite{1996:gibson,2000:gibson,2013:kellman,1996:richman,2017:rau}. Hence, an educational course in hardware reverse engineering should involve repetitive exercises of working with the graphical representations to support students in developing fluency for detecting relevant patterns therein.

\par{\bfseries (4) Instructional Support to Develop Connection-Making Abilities.}
Since gate-level netlists can be represented both graphically and textually, it is important to enable students to develop abilities for connection-making between the two different forms of representation. Students have to learn how text and graphics complement and constrain one another. Since the development of connection-making abilities is based on verbally mediated sensemaking processes \cite{2017:rau}, students should reflect and explain their solutions verbally during lectures.

\section{Educational Research as the Foundation for a Reverse Engineering Lab Course}
\label{ase:section:environment}

We now provide details of the practice-oriented hardware reverse engineering course. Prior to the course and task descriptions in \Section{ase:section:structure}, we first introduce the educational software environment based on the interactive gate-level netlist reverse engineering and manipulation tool \HAL~\cite{tdsc:2018:fyrbiak} (\Section{ase:section:hal}).

\subsection{Educational Software Environment}
\label{ase:section:hal}

\HAL assists users in the reverse engineering of complex gate-level netlists and its extensibility allows for the development of custom plugins. In particular, \HAL employs an interactive \ac{GUI} to provide both textual and graph-based representation of the gate-level netlist under inspection. We stress that a \ac{GUI} is of major importance in manual reverse engineering since making sense of complex or even relatively small hardware circuits consisting of a few hundred gates is considerably easier with a graph-based representation than it is with an inflexible textual representation.

\par{\bfseries Relevance with Respect to Guidelines.}
We want to emphasize that \HAL supports \textit{Guideline 1} as defined in \Section{ase:section:guidelines:hwreg}, e.g. by providing a learning environment, which integrates graphical and textual representations of gate-level netlists.

\subsection{Lab Course Structure and Projects}
\label{ase:section:structure}
Since hardware reverse engineering requires hands-on experience, we decided to offer a 3 ECTS credit point lab course to final year bachelor's and master's students of our IT security programs. Since their relevant prior knowledge does not differ, course requirements are equal for both groups. While the course provides an overview of the multi-layered chip-level reverse engineering processes and hardware security in general, the main learning objectives focus on gate-level netlist reverse engineering using \HAL. In order to assure real-world relevance, we also invited two industry experts who presented specific hardware reverse engineering projects.

\par{\bfseries Structure.}
To facilitate our students' learning success, the course consists of five projects (minimum grade to pass: $75\%$) which have to be completed individually by each participant. Each project lasts three weeks and contains the following sub-tasks: (1)~the reading of relevant scientific papers, (2)~pen~\&~paper exercises, and (3)~practical reverse engineering tasks. The reading and understanding of 1-2 scientific papers conveys relevant content for subsequent tasks while the pen~\&~paper exercises support reproduction and internalization of relevant information. The acquired know\-led\-ge is first applied to small-scale examples, and subsequently in more complex contexts during the practical reverse engineering tasks. At the beginning of each project, theoretical and practical background is taught in one introductory session and after the submission deadline, solutions are discussed and students are encouraged to present their approaches to the class in another session. In total, 12 sessions are held during the 15-week winter term: 10 project sessions and 2 invited sessions.

\par{\bfseries Relevance with Respect to Guidelines.}
The project design supports students as they develop perceptual competencies as defined in \textit{Guideline~3}, e.g. through repeated tasks involving the use of graphical representations.
Furthermore, the course structure satisfies \textit{Guidelines~2} and \textit{4}, e.g. by encouraging active discussions during our lab sessions.
 
\subsubsection{Project 1 -- Standard Cell Reverse Engineering}

To understand the goals and needs of hardware reverse engineering, students first have to acquire general know\-led\-ge of hardware security by answering questions drawn from two comprehensive works~\cite{PotIEEE:2014:rostami,JETC:2016:Quadir} which cover topics ranging from the semiconductor supply chain structure, to hardware design flow, to threat models. Both the pen~\&~paper and the practical exercises focus on standard cell reverse engineering, which is an essential element of chip-level reverse engineering. Once a standard cell is identified, the analyst can employ an image recognition algorithm to find all other occurrences of the same standard cell. This leads to a higher level of abstraction and thus makes it easier to understand the circuit functionality~\cite{JETC:2016:Quadir}.
In the practical assignment, students reverse engineer standard cells from the \ac{IC} layout of an \ac{AES} \ac{IP} core and subsequently utilize \textit{KLayout}~\cite{url:klayout} to automatically extract a gate-level netlist.

\subsubsection{Project 2 -- Netlist Reverse Engineering}

In this exercise, the gate-level netlist analysis and manipulation tool \HAL is introduced to the students.
In addition to \acp{ASIC}, widely-used \acp{FPGA} lend themselves to reverse engineering.
Since \acp{FPGA} enable the students to directly test circuit manipulations as performed in later tasks, we provide the relevant background on the security of \ac{SRAM}-based \acp{FPGA} and outline the major challenges of gate-level netlist reverse engineering~\cite{PotIEEE:2014:trimberger,tect:2013:subramanyan}.
The pen~\&~paper exercise involves the analysis and reverse engineering of \acf{LUT} contents in order to comprehend how combinational circuits are realized on an \ac{FPGA}. In the practical assignment, students must decrypt a given ciphertext, which has been processed by a simple proprietary cryptographic encryption algorithm with a known key. In order to implement the decryption function, students have to analyze the datapath of the encryption algorithm realized by the design under consideration, which is given in the form of a gate-level netlist. More precisely, they have to analyze cryptographic Sboxes (implemented in \acp{LUT}) and assemble the corresponding inverse Sboxes for the decryption function. Sboxes are small tables which are central components of modern ciphers.

\subsubsection{Project 3 -- \acs{FSM} Reverse Engineering}

Since the control logic of most digital systems is implemented by \acfp{FSM}, this project focuses on \acs{FSM} reverse engineering in a realistic scenario. The \acs{FSM} control signals are of particular interest to the reverse engineer because these signals steer the datapath modules of the design or communicate to other \acp{FSM}.
Through the integration of scientific papers~\cite{iscas:2010:shi,aspdac:2016:meade}, we taught students the fundamentals of constructing \acs{FSM} circuits in hardware and the algorithmic identification of \acp{FSM} from gate-level netlists. In the pen~\&~paper assignment, the students have to recover the possible states, draw the state transition graph, and derive further high-level information, e.g. the \acs{FSM} encoding, from a gate-level description of a small \ac{FSM}. An example for such an \acs{FSM} is depicted in \Figure{ase:figure:synthesis_example}.
In the practical assignment, students have to reverse engineer the control logic from a slightly modified variant of the proprietary cipher from Project 2 with the goal of determining the number of iterated rounds which are executed by the cipher. After identifying the combinational logic gates and signals of which the \ac{FSM} is composed, they define the state transition graph and deduce the function of each state.

\subsubsection{Project 4 -- Crypto Reverse Engineering}

While Projects 2 and 3 focus on (a) acquiring know\-led\-ge, (b) learning and implementing basic reverse engineering strategies, and (c) becoming familiar with the handling of \HAL, Project 4 aims at the application of these skills in the context of a real-world implementation of the  \acf{AES} on an \ac{FPGA}. AES is the most widely used encryption algorithm. 
Initially, a brief overview of the \ac{AES} standard and the fundamental concepts of \ac{AES} hardware architectures are provided to the students~\cite{Rodriguez-Henriquez:2010:CAR:1951694}.
Usually, Sboxes are implemented on \acp{FPGA} utilizing \acp{LUT} and multiplexers. In the pen~\&~paper exercise, students reverse engineer a small circuit implementing one column of a 4 bit Sbox. Subsequently, they sketch the hardware implementation of an AES Sbox and calculate the number of potential input permutations for the AES Sbox.
In the practical assignment, students have to extract a hard-coded key from a real-world AES core to decrypt a given ciphertext. After deriving high-level information like the functionality (encryption or decryption), the presence of the key schedule, the key length, and the hardware architecture (iterative or pipelined), they have to write a \HAL plugin to identify the Sbox logic, since the Sboxes serve as a potential anchor for attacks on the hard-coded key. Finally, the hard-coded key is extracted through manipulation of the underlying circuit (insertion of a hardware Trojan) or through exhaustive reverse engineering.

\subsubsection{Project 5 -- Reverse Engineering vs. Obfuscation}

This project introduces hardware obfuscation as a potential countermeasure to IP theft and reverse engineering to the students~\cite{book:hardware_obfuscation:chapter1}.
Control flow obfuscation methods are discussed as a specific example of obfuscation methods on the netlist-level.
The reading briefly introduces a weak \acs{FSM}-based obfuscation method utilizing so-called obfuscated states that have to be passed in correct order before the original states appear~\cite{iccad:2009:chakraborty}.
In the combined pen~\&~paper and practical assignments, students defeat the introduced obfuscation method through using structural reverse engineering and manipulation of the gate-level netlist.
Students are equipped with a gate-level netlist and the corresponding bitstream that is implementing the aforementioned obfuscation scheme.
While the bitstream implements the obfuscated \acs{FSM} on a small \acs{FPGA}, the students can control the \acs{FSM} directly via UART, and hence test the successfulness of their attempts by receiving a visual feedback from an LED.
To accomplish this task, students have to manipulate the gate-level netlist file with \HAL and synthesize it to the proprietary bitstream using the vendor toolchain.

\section{Evaluation -- Methods and Results}
\label{ase:section:evaluation}
\subsection{Methods}
To the best of our knowledge, the proposed course is the first to teach domain-specific knowledge in hardware reverse engineering with a particular focus on gate-level netlist reverse engineering incorporating educational theory. One already existing course for teaching hardware reverse engineering \cite{url:zonenberg} was developed and conducted by Yener and Zonenberg, but did not focus on gate-level netlist reverse engineering in particular.

Because of our unique focus on teaching gate-level netlist reverse engineering we could not draw much educational guidance from existing courses. This resulted in the need to perform a thorough course evaluation to gain deeper insights for future lectures in this field. Overall, 10 students participated in the pilot course. All of them were enrolled in IT Security programs at Ruhr-Universität Bochum, Germany. Five students (all male; mean age 22.2 years; 3 on bachelor's level; 2 on master's level) consented to the analysis of their data within the context of this educational evaluation.

The main goals of our evaluation were: to understand how challenging the projects were and to what extent their difficulty was dependent upon the students' prior level of knowledge, to assess whether the workload corresponding to each individual project was adequate, to determine if \HAL and its graphical representations supported the learning process, and to identify what other difficulties, if any, the students encountered. In the following, we introduce the evaluative methods we used, summarize the results, and highlight implications for future courses on hardware reverse engineering.

\subsubsection{Relevant Prior Knowledge}
We designed our projects and introductory lectures based upon assumptions about the students' level of prior knowledge of relevant topics. By measuring their level of prior knowledge, we were able to detect knowledge gaps which explain difficulties they encountered and help us revise the course for future iterations. We used two methods to measure the level of relevant prior knowledge. The first method consisted of three self-developed items including the following questions: \enquote{\itshape How highly do you rate your level of prior knowledge in hardware reverse engineering?} Students had been asked to rate their answer on a 5-point \textit{Likert Scale}~\cite{1932:likert}, ranging from \enquote{\itshape very low (1)} to \enquote{\itshape very high (5)}. The other two items \enquote{\itshape How much theoretical exposure to hardware reverse engineering have you had?} and \enquote{\itshape How much practical experience in hardware reverse engineering do you have?} had to be answered on a 5-point Likert scale~\cite{1932:likert}, ranging from \enquote{\itshape very much (1)} to \enquote{\itshape not at all (5)}. Additionally, students were asked to provide information about any relevant courses they successfully completed.

\subsubsection{Cognitive Load -- Mental Effort and Perceived Task Difficulty}
Besides the level of prior knowledge, we measured mental effort and perceived task difficulty. The methods used in this study are common in current research on cognitive load~\cite{2015:schmeck} by means of two subjective rating scales.
The first scale was the mental effort rating scale~\cite{1992:paas}. We asked students to rate their invested amount of mental effort on a 7-point \textit{Likert Scale}~\cite{1932:likert}, ranging from \enquote{\itshape very low mental effort (1)} to \enquote{\itshape very high mental effort (7)} \cite{2015:schmeck}. Another commonly used scale to measure cognitive load is the perceived task difficulty scale~\cite{1972:bratfisch, 1999:kalyuga, 1996:marcus, 2003:paas}. Students were asked to rate the perceived task difficulty on a 7-point \textit{Likert Scale}~\cite{1932:likert}, ranging from \enquote{\itshape very easy (1)} to \enquote{\itshape very difficult (7)}.

\subsection{Results}
In summary, all 10 students completed the course successfully. In the following, we present and discuss the results of the 5 students who consented to evaluation. Here we would like to note that we did not attempt to determine the probability of successful course completion in our analysis. As stated before, students had to reach at least 75\% to pass a single project and all 10 students reached between 80-100\% in each project. Based on this structure, the probability of successful course completion has no explanatory power.

\begin{table*}[htb!]
	\centering
	\caption{Descriptive analysis of prior knowledge with means and standard deviations}
	\label{ase:table:experiences}
	\begin{tabular}{lccc}
		\Xhline{2\arrayrulewidth}
		& Prior Knowledge & Theoretical Experiences & Practical Experiences\\
		\hline
		Mean (SD)  & 2.00 (1.00)    & 4.20 (.84)&   4.60 (.55)\\
		\Xhline{2\arrayrulewidth}
	\end{tabular}
\end{table*}
\begin{table}[htb!]
	\centering
	\caption{Descriptive analysis of perceived task difficulty and mental effort with means and standard deviations}
	\label{ase:table:cognitive_load}
	\begin{tabular}{ccc}
		\Xhline{2\arrayrulewidth}
		& Perceived Task Difficulty & Mental Effort\\
		\hline
		Project 3   & 3.60 (1.14)    & 4.20 (.84)\\
		
		Project 4   & 3.60 (.89)    & 4.40 (.85)\\
		
		Project 5   & 5.20 (.45)    & 5.60 (.89)\\
		\Xhline{2\arrayrulewidth}
	\end{tabular}
\end{table}

To set the particular focus of the evaluation on the novel hardware reverse engineering tasks, students were asked to rate projects 3-5. Those three projects included real-world hardware reverse engineering tasks with growing complexity. Note that project 1 and 2 were important to introduce hardware reverse engineering in general and the educational software HAL in particular, but did not include real-world hardware reverse engineering tasks. Thus, they were excluded from the evaluation of this course.

\par{\bfseries Level of Relevant Prior Knowledge.}
Most of the students had low theoretical knowledge and barely any practical experience in hardware reverse engineering. \Table{ase:table:experiences} illustrates the results of a descriptive analysis with means and standard deviation (SD). In addition, most of the students had not completed relevant coursework and consequently had gaps in domain-specific topics like the application of Boolean functions or the implementation of cryptographic schemes. 

\par{\bfseries Level of Mental Effort and Task Difficulty.}
\Table{ase:table:cognitive_load} summarizes the descriptive analysis of cognitive load. Means and standard deviations revealed moderate levels of perceived task difficulty and mental effort.


\section{Discussion and Implications}
\label{ase:section:discussion}

All 10 students successfully passed the pilot course. First, this result supports the assumption that the course and its foundation based on the educational guidelines contribute to teaching hardware reverse engineering successfully. Project~5 included the most complex task which is reflected in higher means of perceived task difficulty and mental effort.
Although the task complexity gradually increases from project 3 to project 5, all students successfully passed each single task which leads to the inference that the course structure contributes to learning hardware reverse engineering efficiently.

 Second, the combination of graphical and textual representations as integrated in the course supported the students' learning success by making abstract concepts in hardware reverse engineering more accessible. Third, the course structure and the inclusion of instructional support facilitated the acquisition of relevant competencies and connection-making abilities, which are necessary to learn from and work with graphical and textual representations as used in HAL. 

\par{\bfseries Course Restructuring -- Closing Prior Knowledge Gaps.}
Despite the successful implementation of the pilot course, the challenges presented by the students' lack of relevant knowledge prior to beginning the course merits further discussion. Through their feedback, we determined that the students had knowledge gaps in relevant content area, which caused them to invest more time than we had intended to complete the projects. To provide greater support for students’ learning processes, we decided to partially restructure the material of our course in the future. Instead of a single practice-oriented lab course, we are currently designing a new series which includes fundamental background on the hardware design processes in Lectures 1-6 (from~\cite{Weste:2010:CVD:1841628}). From there, Lectures 7-12 will holistically capture the multi-layer hardware reverse engineering process (e.g., from~\cite{dt:1999:hansen,PotIEEE:2014:rostami,JETC:2016:Quadir,ches:2009:torrance,PotIEEE:2014:trimberger,ivsw:2017:fyrbiak}). We also set an appropriate incentive that reflects the expected workload by offering 5 ECTS credit points for successful participation in the lectures, in addition to 3 ECTS for the lab part of the course.

To test the course structure and to obtain deeper insights into the learning processes of a broader student population, we plan to offer the modified course for the first time at the University of Massachusetts Amherst, USA, and Ruhr-Universität Bochum, Germany, in the first six weeks (2 lectures per week) of the 2018/19 winter term. In the last 9 weeks of the semester, we will offer our lab course. We expect our students to satisfy the shorter project time limits (2 instead of 3 weeks) thanks to their significantly higher prior knowledge in relevant topics.

\section{Conclusion}
\label{ase:section:conclusion}

In this paper, we highlighted the necessity of security engineers' and system designers' acquisition of solid hardware reverse engineering skills. Even though there is an increasing demand for specialists in this area, we identified a lack of educational courses which make hardware reverse engineering more accessible to students by teaching fundamental background content in a well-structured and practice-oriented fashion.

To this end, we formulated course structure guidelines based on insights gleaned from the fields of educational research and cognitive science which led to the development of a novel course on hardware reverse engineering with particular focus on gate-level netlist reverse engineering. Based on the educational evaluation of our first pilot lab course, we identified important aspects of future improvements to course content, particularly the incorporation of fundamental background knowledge of hardware design processes, an area in which students were lacking.

\bibliographystyle{IEEEtr}
{
	\footnotesize \bibliography{IEEEabrv,bibliography}
}

\end{document}